\documentclass{PoS}
\usepackage{epsfig}

\title{Femtoscopy of proton-proton collisions at the LHC with the ALICE experiment}

\ShortTitle{Femtoscopy of pp collisions in ALICE}

\author{\speaker{{\L}ukasz Graczykowski} (for the ALICE Collaboration)\\
        Warsaw University of Technology, Poland\\
        E-mail: \email{lgraczyk@if.pw.edu.pl}}

\abstract{We present the results on two-particle Bose-Einstein correlations measured in proton-proton collisions at center of mass energies of $\sqrt{s}=0.9\ \rm{TeV}$, $\sqrt{s}=2.76\ \rm{TeV}$ and $\sqrt{s}=7\ \rm{TeV}$ registered by the ALICE experiment at the Large Hadron Collider. Detailed analysis reveals that the three dimensional experimental correlation functions do not have Gaussian shape in two dimensions. We found that they are better described by an exponential functional form in the outward and longitudinal directions, while the sideward remains a Gaussian. This is interpreted as a result of a significant contribution of strongly decaying resonances to the shape of the emission region.}

\FullConference{The Seventh Workshop on Particle Correlations and Femtoscopy\\
		 September 20 - 24 2011\\
		 University of Tokyo, Japan}

\begin{document}

\section{Introduction}
The Large Hadron Collider (LHC) at CERN started operating in the end of the year 2009 and has been colliding protons at the center of mass energies of $\sqrt{s}=0.9\ \rm{TeV}$, $\sqrt{s}=2.76\ \rm{TeV}$ and $\sqrt{s}=7\ \rm{TeV}$. At the LHC, the ALICE experiment \cite{ALICE} focuses mainly on measuring heavy-ion collisions. However, minimum-bias proton-proton collisions which provide the heavy-ion 'baseline' are also measured. The proton-proton data has been used by the ALICE \cite{FemtoPaper,NewFemtoPaper} and CMS \cite{CMSpaper} collaborations in the two-pion Bose-Einstein femtoscopic analyses. In the detailed studies performed by ALICE on $0.9\ \rm{TeV}$ and $7\ \rm{TeV}$ data, in three dimensions, we showed that the shape of the experimental correlation function is clearly not Gaussian \cite{NewFemtoPaper}. We tested different functional forms in each \emph{out}, \emph{side} and \emph{long} directions and proposed a better fitting formula according to the best fit. In this work we extend the analysis to $\sqrt{s}=2.76\ \rm{TeV}$ data.

\section{Data analysis}
Data samples of approximately 4 million proton-proton collisions at center of mass energies of $\sqrt{s}=0.9\ \rm{TeV}$, 20 million at $\sqrt{s}=2.76\ \rm{TeV}$ and 60 million at $\sqrt{s}=7\ \rm{TeV}$ were analyzed. The subsystems used for the analysis were the Inner Tracking System (ITS), the Time Projection Chamber (TPC) and the VZERO detectors. All the events were required to have a reconstructed interaction point (primary vertex) within 10 cm from the center of the ALICE detector in the beam direction. The ITS and TPC were used for tracking in the pseudorapidity range $|\eta|<1.0$ while the VZERO was used for a minimum bias trigger and to reject beam-gas and beam-halo collisions. Identification of pions was based on the energy loss of the particle information from the TPC. The analysis was always performed on primary particles and their selection was based on the minimum distance between the track and the primary vertex (so-called \emph{Distance of Closest Approach} or DCA). Tracks were required to have DCA not greater than $0.018+0.035p_T^{-1.01}$ in the transverse plane and $0.3\ \rm{cm}$ in the longitudinal direction. We also applied specific procedures to suppress undesired two-track effects, such as splitting (one track reconstructed as two) and merging (two tracks reconstructed as one). For more details about the event and particle selection criteria see \cite{NewFemtoPaper}.
\\The analysis was performed in 8 ranges of total measured charged-particle multiplicity of the event $N_{ch}$ and 6 ranges of pair transverse momentum $k_T=\frac{|\vec{p_1}+\vec{p_2}|}{2}$.

\section{Results of the femtoscopy analysis}
\begin{figure}[!ht]
\centering
\epsfig{file=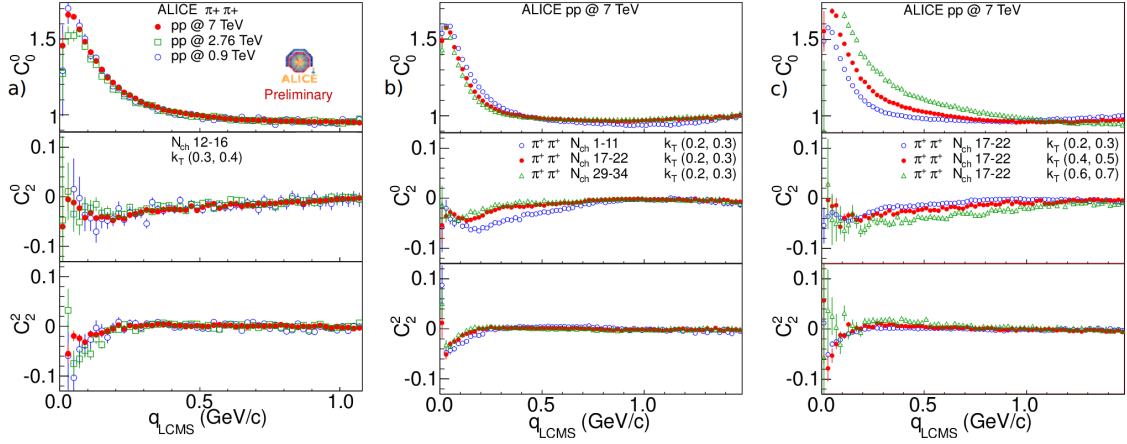,width=15 cm}
\caption{\emph{Comparison of the correlation functions for $\pi^+\pi^+$ for a) different collision energies, b) different multiplicity ranges, c) different pair transverse momentum ranges.}}
\label{fig:CFs}
\end{figure}

The femtoscopic analysis was performed in order to obtain the sizes of the particle emitting region. The Spherical Harmonics decomposition of the three-dimensional correlation functions was applied. This technique allows to represent the three-dimensional object as an infinite set of one-dimensional spherical harmonics functions.
The symmetries of the pair distribution make most of the components vanishing and the first three of the non-zero ones, $C_0^0$, $C_2^0$ and $C_2^2$, capture most important information about the correlation \cite{ChajeckiLisa,KisielBrown}. The first one is the angle-averaged component, the second measures the difference between \emph{out} and \emph{side} while the third is the difference between \emph{transverse} and \emph{long}.
The results show that the correlation functions are practically independent on the collision energy but they depend on the event multiplicity and (more strongly) on pair transverse momentum. Increasing the collision energy by an order of magnitude has less impact on the experimental correlation functions than changing the multiplicity by 50\%. All these effects are presented in Fig. \ref{fig:CFs}.

Let us focus on the $C_0^0$ component. The femtoscopic effect, coming from the symmetrization of the two-pion wave function, is visible as the increase of the correlation function below $q=0.5\ \rm{GeV/c}$. The fall visible in the lowest bins is due to the Coulomb repulsion, but it has little impact on the Bose-Einstein peak and, therefore, on the extracted radii. We also see a broad peak extending in $q$ from 0 up to, at least, $1.0\ \rm{GeV/c}$. We performed a very detailed analysis of this effect in \cite{NewFemtoPaper}. Because such structure can be seen in Monte Carlo data, which does not include Bose-Einstein correlations, it has clearly non-femtoscopic origin. Therefore, we can try to parameterize it using Monte Carlo simulations and use this parameterization to fit the data. From various formulas that we had tried to apply we found that it is quite well described by a Gaussian form in $C_0^0$ and $C_2^2$. Therefore, the final functional form for the background treatment is:
\begin{equation}
B(\mathbf{q})=A_h\exp\left ( -q^2A_w^2\right ) + B_h\exp\left ( \frac{-(q-B_m)^2}{2B_w^2}  \right )(3\cos^2(\theta)-1).
\label{eq:corrbkg}
\end{equation}
The analysis of the minijet origin of the non-femtoscopic background using angular correlations is described in \cite{majanik}.

\section{Fitting of the correlation function}
\subsection{Gaussian fits}
The femtoscopic correlation function is defined by the Koonin-Pratt formula \cite{Koonin}:
\begin{equation}
C(\vec{q})=\int S(\mathbf{r})|\Psi(\mathbf{r},\vec{q})|d^4\mathbf{r}.
\label{eq:KooninPratt}
\end{equation}
Usually it is assumed that the emission function describes a static source and has an ellipsoid Gaussian profile in space:
\begin{equation}
S(\mathbf{r})=S(r_{o},r_{s},r_{l})\sim \exp\left ( -\frac{r_{o}^{2}}{4R_{out}^{2}}-\frac{r_{s}^{2}}{4R_{side}^{2}}-\frac{r_{l}^{2}}{4R_{out}^{2}} \right ),
\end{equation}
where $R_{i}$ are the sizes of the source and $\mathbf{r}=[r_{out},r_{side},r_{long}]$ is the pair separation vector.
Using formula (\ref{eq:KooninPratt}) and such parametrization of the emission function, one gets the following form of the correlation function:
\begin{equation}
\label{eq:corrfctnLCMS}
C(\mathbf{q})=C(q_{out},q_{side},q_{long})=1+\lambda\exp\left (-R_{out}^{2}q_{out}^{2}-R_{side}^{2}q_{side}^{2}-R_{long}^{2}q_{long}^{2} \right ),
\end{equation}
where $\lambda$ is the strength of the correlation (fraction of correlated pairs for which both pions were correctly identified).
In the data analysis we must also take into account the Coulomb interaction between identical pions and strong interactions in the final state. However, since pions are emitted from sources which are on the order of $2-3\ \rm{fm}$ in size, the strong contribution is small and can be neglected \cite{Lednicky2009}. The influence of Coulomb interaction is approximated using the \emph{Bowler-Sinyukov formula} which assumes that the Coulomb part can be factorized out from the wave function of the pair and integrated separately. The modified correlation function, including Coulomb interaction, is given by the equation:
\begin{equation}
C_{f}(\mathbf{q})=(1-\lambda)+\lambda K(\mathbf{q})\cdot[1+\lambda\exp\left (-R_{out}^{2}q_{out}^{2}-R_{side}^{2}q_{side}^{2}-R_{long}^{2}q_{long}^{2} \right )],
\end{equation}
where $K$ is the Coulomb like-sign pion pair wave function averaged over the Gaussian source with a radius of $1\ \rm{fm}$.
Because the Monte Carlo generators do not include Bose-Einstein correlations, they can be used to estimate the non-femtoscopic background of the correlation function. In order to do this we fit the Monte Carlo simulated data with equation (\ref{eq:corrbkg}) described in the previous section. The final form of the fitting function is therefore:
\begin{equation}
C(\mathbf{q})=C(q_{out},q_{side},q_{long})=NC_{f}(q_{out},q_{side},q_{long})B(q_{out},q_{side},q_{long}),
\label{eq:corrfit}
\end{equation}
where $N$ is the overall normalization.
The Gaussian fits to the experimental correlation functions are presented in the left panel of Fig. \ref{fig:Fits}. It clearly underestimates the height of the peak. However the width of the correlation function is reproduced.

\subsection{Non-Gaussian fits}
\begin{figure}[!ht]
\centering
\epsfig{file=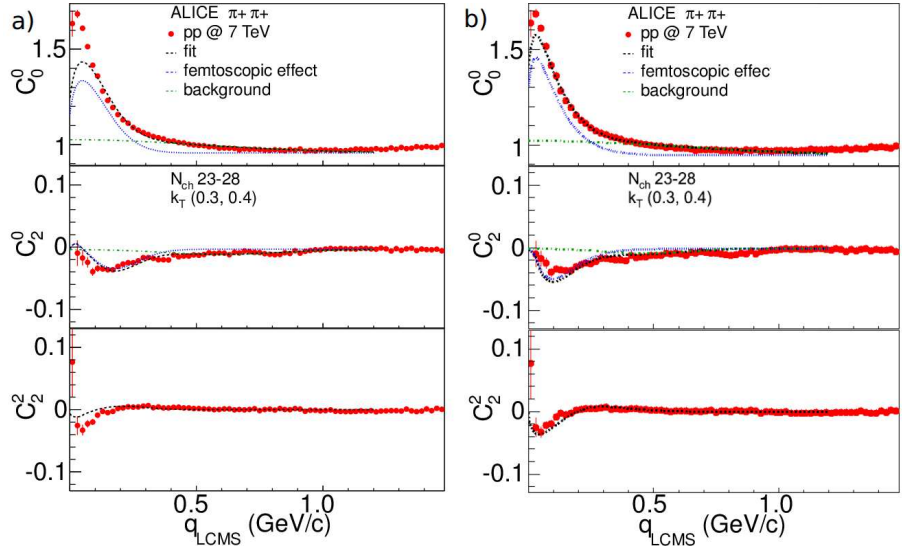,height=7.205 cm}
\caption{\emph{Gaussian (a) and non-Gaussian (b) fits to the correlation function for the $\sqrt{s}=7\ \rm{TeV}$ collisions with $23<N_{ch}<29$ and pairs with $0.3<k_T<0.4\ \rm{GeV/c}$.}}
\label{fig:Fits}
\end{figure}

In the previous section we have shown that the Gaussian fits do not perfectly describe the correlation functions obtained from the experimental data. In order to improve the fit we assume that the source function factorizes into \emph{out}, \emph{side} and \emph{long} directions:
\begin{equation}
S(\mathbf{r})=S(r_{out})S(r_{side})S(r_{long}).
\label{eq:sourfact}
\end{equation}
This leads to the factorization of the correlation function itself:
\begin{equation}
C(\vec{q})=1+\lambda C(q_{out})C(q_{side})C(q_{long}).
\label{eq:corrfctnfact}
\end{equation}
We can change the functional form of each of the components of the source (and therefore the correlation function) independently. Three different functional forms of the source function were analyzed - Gaussian, exponential and Lorentzian. They have the desired feature, that the integration in formula (\ref{eq:KooninPratt}) can be performed analytically and lead to the Gaussian, Lorentzian and exponential forms of the correlation function respectively. It is reasonable to fit the data with functional forms other than Gaussian, especially in \emph{out} and \emph{long} directions, because resonances decay after random time and this process is determined by the exponential decay law, which transforms into an exponential shape in space via the pair velocity (which by definition exists in \emph{out} and \emph{long} directions, and vanishes in \emph{side} direction). We performed a study of all 27 combinations of the fitting functions for all the multiplicity and $k_T$ ranges. We found that universally in \emph{out} direction the correlation function was best described by an exponential form, corresponding to Lorentzian emission function, which agrees with model expectations. In contrast, the \emph{side} direction is equally well described by a Gaussian or a Lorentzian: we chose the former because the Lorentzian correlation function would correspond to exponential pair emission function with a sharp peak at 0.
\\The non-Gaussian fit is shown in the right panel of Fig. \ref{fig:Fits}.
One can see that the exponential form in \emph{out} and \emph{long} reproduces the data much better, for both the peak height and its width.

\subsection{Radii dependencies}
\begin{figure}[!ht]
\centering
\epsfig{file=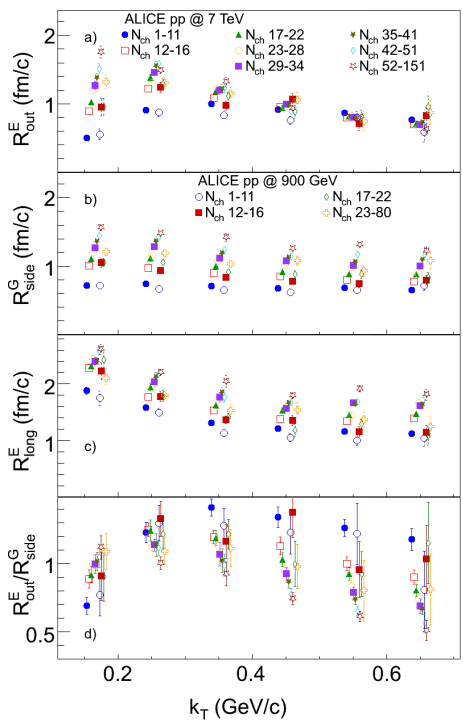,height=9.6 cm}
\caption{\emph{Pair transverse momentum and multiplicity dependencies of the femtoscopic radii from the non-Gaussian fits to the correlation functions from $7\ \rm{TeV}$ data.}}
\label{fig:Radii}
\end{figure}

The dependencies of the femtoscopic radii on event multiplicity and $k_T$ extracted from non-Gaussian fits are shown in Fig. \ref{fig:Radii}. We stress that the radii cannot be directly compared to the ones from the Gaussian fits. We see that $R_{long}$ and $R_{side}$ always fall with $k_T$. On the other hand, $R_{out}$ has a different behavior - it firstly rises for the lowest $k_T$ ranges and then goes down. The radii also scale linearly with multiplicity. This scaling is observed in every $k_T$ range. $R_{long}$ and $R_{out}$ radii always grow with multiplicity while $R_{out}$ radius grows for the lower $k_T$ range and falls for the higher $k_T$ range. Gaussian fits show similar behavior.

\section{Conclusions}
We performed a detailed study of the Bose-Einstein femtoscopic correlations in proton-proton collisions at center of mass energies of $\sqrt{s}=0.9\ \rm{TeV}$, $\sqrt{s}=2.76\ \rm{TeV}$, $\sqrt{s}=7\ \rm{TeV}$ collected by the ALICE experiment. We found that the correlation functions are almost independent on collision energy, however they depend on the event multiplicity and on the pair transverse momentum $k_T$. The experimental correlation functions are clearly not Gaussians. We introduced an assumption that the source function factorizes, which leads to the factorization of the correlation function. Each component of the correlation function can be described by either Gaussian, exponential or Lorentzian functional form in $out$, $side$ and $long$ directions. From the fitting of all possible combinations in all multiplicity and $k_T$ ranges we found that the exponential form in $out$ and $long$ and Gaussian form in $side$ fit data best. The extracted radii from both Gaussian and non-Gaussian fits show similar behavior.

\end{document}